\documentclass[conference]{IEEEtran}
\IEEEoverridecommandlockouts
\usepackage{cite}
\usepackage{amsmath,amssymb,amsfonts}
\usepackage{algorithmic}
\usepackage{graphicx}
\usepackage{textcomp}
\usepackage{xcolor}
\usepackage[linesnumbered,ruled,vlined]{algorithm2e}
\usepackage{amsmath,amsfonts}
\usepackage{array}
\usepackage[caption=false,font=normalsize,labelfont=sf,textfont=sf]{subfig}
\usepackage{textcomp}
\usepackage{stfloats}
\usepackage{url}
\usepackage{verbatim}
\usepackage{graphicx}
\usepackage{cite}
\usepackage{pifont}
\usepackage[linesnumbered,ruled,vlined]{algorithm2e}
\usepackage{color}
\usepackage{booktabs}
\usepackage{tabu}
\usepackage{multirow}
\usepackage{graphicx}
\usepackage{float}
\usepackage{subfig}
\usepackage[misc]{ifsym}
\def\BibTeX{{\rm B\kern-.05em{\sc i\kern-.025em b}\kern-.08em
    T\kern-.1667em\lower.7ex\hbox{E}\kern-.125emX}}
\begin{document}

\title{Listen to Minority: Encrypted Traffic Classification for Class Imbalance with Contrastive Pre-Training
}

\author{
  \IEEEauthorblockN{
    Xiang~Li\IEEEauthorrefmark{1}\IEEEauthorrefmark{2},
    Juncheng~Guo\IEEEauthorrefmark{3},
    Qige~Song\IEEEauthorrefmark{4},
    Jiang~Xie\IEEEauthorrefmark{4},
    Yafei~Sang\IEEEauthorrefmark{1},
    Shuyuan~Zhao\IEEEauthorrefmark{1}$^{(\textrm{\Letter})}$,
    and Yongzheng~Zhang\IEEEauthorrefmark{5}
  }
  \IEEEauthorblockA{\IEEEauthorrefmark{1} Institute of Information Engineering, Chinese Academy of Sciences, Beijing, China}
  \IEEEauthorblockA{\IEEEauthorrefmark{2} School of Cyber Security, University of Chinese Academy of Sciences, Beijing, China}
    \IEEEauthorblockA{\IEEEauthorrefmark{3} Amazon, China}
    \IEEEauthorblockA{\IEEEauthorrefmark{4}  Zhongguancun Laboratory, Beijing, China}    
   
  \IEEEauthorblockA{\IEEEauthorrefmark{5} China Assets Cybersecurity Technology CO.,Ltd., Beijing, China}
  {Email: \{lixiang1, zhaoshuyuan\}@iie.ac.cn}
}
\maketitle

\begin{abstract}
Mobile Internet has profoundly reshaped modern lifestyles in various aspects. Encrypted Traffic Classification (ETC) naturally plays a crucial role in managing mobile Internet, especially with the explosive growth of mobile apps using encrypted communication. Despite some existing learning-based ETC methods showing promising results, three-fold limitations still remain in real-world network environments, \romannumeral1) label bias caused by traffic class imbalance, \romannumeral2) traffic homogeneity caused by component sharing, and \romannumeral3) training with reliance on sufficient labeled traffic. None of the existing ETC methods can address all these limitations. In this paper, we propose a novel Pre-trAining Semi-Supervised ETC framework, dubbed PASS. Our key insight is to resample the original train dataset and perform contrastive pre-training without using individual app labels directly to avoid label bias issues caused by class imbalance, while obtaining a robust feature representation to differentiate overlapping homogeneous traffic by pulling positive traffic pairs closer and pushing negative pairs away. Meanwhile, PASS designs a semi-supervised optimization strategy based on pseudo-label iteration and dynamic loss weighting algorithms in order to effectively utilize massive unlabeled traffic data and alleviate manual train dataset annotation workload. PASS outperforms state-of-the-art ETC methods and generic sampling approaches on four public datasets with significant class imbalance and traffic homogeneity, remarkably pushing the F1 of Cross-Platform215 with 1.31\%$\uparrow$, ISCX-17 with 9.12\%$\uparrow$. Furthermore, we validate the generality of the contrastive pre-training and pseudo-label iteration components of PASS, which can adaptively benefit ETC methods with diverse feature extractors. 
\end{abstract}


\section{Introduction}
With the unprecedented blossoming of the mobile Internet, ubiquitous mobile apps have become necessities for the population, profoundly reshaping various aspects of modern lifestyles. For mobile Internet management, network operators need to understand and classify high volume of traffic and correlate them with the mobile apps they generate, which is the backbone of network monitoring services as a critical prerequisite of QoS \cite{zhao2021network}. Mobile apps provide users with high-quality services, accompanied by collecting and transmitting user data, raising significant privacy concerns. To alleviate this problem, mobile app developers adopt TLS encryption technology to transmit plaintext user data, which in turn brings thorny challenges to mobile Internet management, defeating traditional deep packet inspection-based and port-based traffic classification methods \cite{letsencrypt}. 

Researchers have proposed solutions for Encrypted Traffic Classification (ETC) by extracting classifiable side-channel information without inspecting packet payload, thus avoiding the impact of encryption techniques \cite{wang2017malware, liu2019fs, zhang2019pccn, he2022payload}. However, existing ETC methods face three-fold limitations:

\textbf{Label bias caused by traffic class imbalance.}
Although several ETC methods seem to produce decent results, most are based on ideal balanced datasets, ignoring that traffic classes are often wildly unbalanced in real-world network environments. According to the QuestMobile report \cite{china2021insight}, 
the traffic of mainstream app families, \emph{e.g.}, Alibaba and Tencent, accounts for more than 66.9\% of the Chinese app market. In comparison, the largest share of other apps is less than 8\%. This phenomenon is more severe in the malware traffic classification scenario, where normal traffic is always much more than malicious traffic. Even only for ransomware, different family classes also show an uneven distribution of more than 20:1 \cite{chen2018machine}. Label bias is one of the problems most directly caused by traffic class imbalance, \emph{i.e.}, the classifier tends to divide the minority class into the majority. Therefore, a robust ETC model that can handle class imbalance is needed.

\textbf{Traffic homogeneity caused by component sharing.}
In real-world scenarios, different apps inevitably share similar network traffic characteristics, resulting in severe traffic homogeneity. The reason is that many app developers use common repositories and associated domains to implement authentication, advertising, and analytics functions. Additionally, some app content is served through the same cloud provider and CDN \cite{yun2022encrypted}. For example, when the  \emph{Taobao} app runs, not all traffic is directly related to Taobao itself. Some actions lead to \emph{amap.com} or \emph{alicdn.com}, which may also appear when running \emph{Alipay} or \emph{Youku} from the Alibaba family. The unclear classification boundary of homogeneous traffic exacerbates the label bias issue, resulting in increased misclassification.  There is still a lack of research on homogeneous traffic, which will help enhance the performance of ETC methods.

\textbf{Training with reliance on sufficient labeled traffic.}
Most ETC methods rely on sufficient training data with explicit class annotation information. Correctly annotating massive traffic is still challenging. Researchers usually collect the labeled traffic of single apps and a few background traffic in restricted networks to ensure correctness of annotations\cite{li2022foap}. In contrast, unlabeled traffic is always easy to obtain and can reflect the data distribution in real-world network environments. A large amount of unlabeled data is considered for use, which can effectively improve the performance of ETC tasks.

In this paper, we propose \textbf{PASS}, a novel \textbf{P}re-tr\textbf{A}ined \textbf{S}emi-\textbf{S}upervised model for ETC, to address the above limitations. 
As the basis of the PASS framework, a multi-granularity traffic sequence construction and a multi-head attention encoder are used to obtain a robust traffic feature representation. To solve the performance degradation caused by traffic class imbalance and homogeneity problems, PASS adopts the idea of contrastive pre-training. Specifically, we design a training data resampling algorithm to construct individual traffic samples into positive and strong/weak negative sample pairs based on their class grouping and communication information. We then optimize the encoder by pushing the positive pairs closer and the negative away to build the traffic feature discrimination capabilities. 
In order to effectively exploit massive unlabeled encrypted traffic data in the network environment, we use the encoder with a classification layer, fine-tuned by original training data, to annotate the unlabeled traffic as pseudo-labeled data, which then is determined the availability according to an appropriate threshold obtained by cross-validation. We set up a weight sampling strategy and a dynamic loss function to balance the contributions of original training data and pseudo-labeled data. 
The above operations bring three salient advantages, \romannumeral1) it can significantly expand existing training data and avoid directly using class labels, which improves the learning ability of the encoder for minority app characteristics to alleviate the label bias problem caused by traffic class imbalance; \romannumeral2) it can help the encoder notice distinguishing overlapping features of homogeneous traffic, by constructing weak negative pairs with communication information to present more fine-grained alignment and uniformity in high-dimensional space; \romannumeral3) it can reduce the dependence on the annotated traffic training data, further improving the accuracy and generalization of ETC methods in real-world scenarios.

\textbf{Our major contributions can be summarized as follows:}

\begin{itemize}
    \item We propose a new ETC working paradigm, which performs contrastive pre-training on the basis of multi-granularity traffic sequence construction and multi-head self-attention encoder, to benefit the ETC task in real-world network environments. To the best of our knowledge, PASS is the first work to address the class imbalance and traffic homogeneity challenges simultaneously. 
    \item We leverage pseudo-label iteration and dynamic loss weighting methods to perform semi-supervised optimization. In this way, PASS can effectively utilize massive unlabeled traffic data in the network environment, alleviating the dependence on labeled training data.
    \item We conduct extensive experiments on four datasets with significant class imbalance and traffic homogeneity problems. Results demonstrate that PASS can surpass state-of-the-art ETC methods and is superior to generic over-sampling methods in addressing the class imbalance. Moreover, the contrastive pre-training and pseudo-label iteration strategies of PASS can be versatilely applied to ETC methods with different traffic feature extractors and adaptively benefit their performance.
\end{itemize}

\section{Overview}
\subsection{Threat Model}
Adversaries can use ETC to perform traffic analysis attacks that identify the number of times a victim has accessed a specific set of monitored applications.
We assume that the potential adversary is local and passive, \emph{i.e.}, the adversary exists in the local network of victims and only performs traffic collection, not active attacks such as traffic hijacking. We define an encrypted network flow as a bidirectional sequence of packets corresponding to a socket-to-socket communication identified by a unique five-tuple \{\emph{source IP, destination IP, source port, destination port, protocol}\}. We focus on TCP flows and consider encrypted traffic based on TCP protocol (\emph{e.g.}, HTTPS). Meanwhile, we assume that the adversary \emph{cannot exploit} the packet plaintext payload and the destination feature, in which the plaintext payload cannot be accessed in the case of encryption, and widespread availability of homogeneous traffic leads to sharing destination features, respectively.

\subsection{Workflow of PASS}
\begin{figure*}[t]
\centering
\includegraphics[scale=0.6]{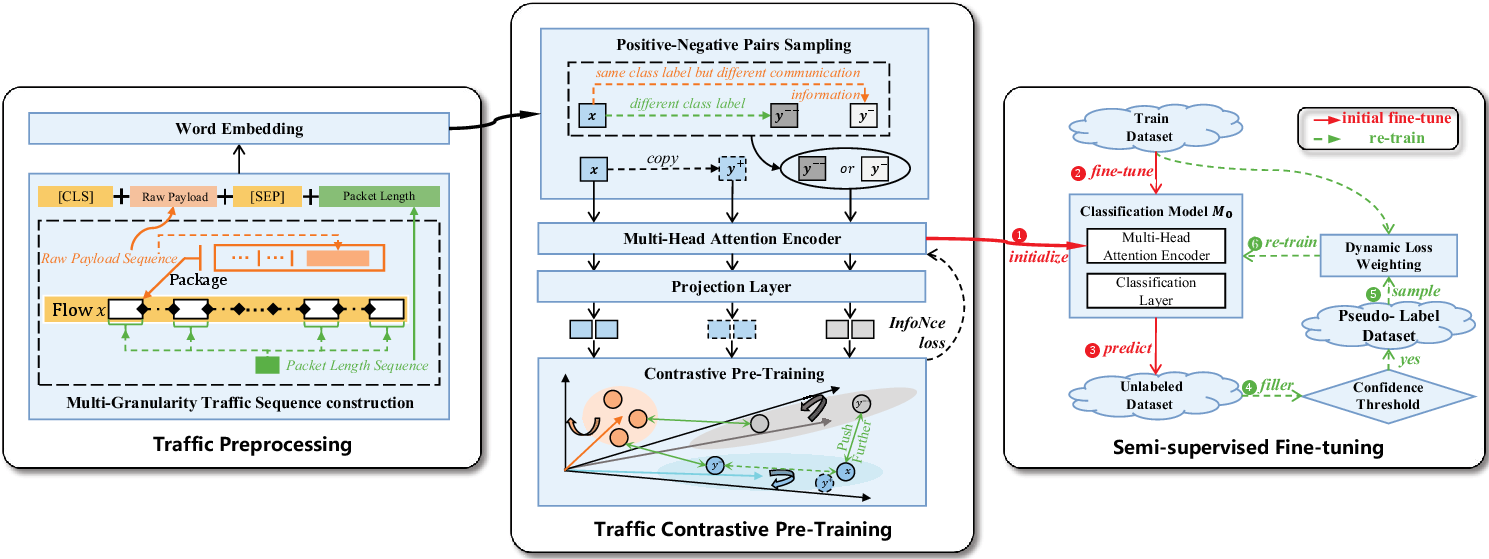}
\caption{Overview of PASS Workflow.}
\label{fig1}
\end{figure*}

Without loss of generality, we assume $x$ is one of the bidirectional flows of interest. Fig.\ref{fig1} illustrates the workflow of PASS, including three stages of \emph{traffic preprocessing}, \emph{traffic contrastive pre-training}, and \emph{semi-supervised fine-tuning}. 

\subsubsection{Traffic Preprocessing}
We extract the Raw Payload (RP) sequence of $m$ bytes and Packet Length (PL) sequence of $n$ packets, reconstruct and encode their appearing elements into a coding dictionary by order of frequency of occurrence. Unique markers CLS and SEP are added, which denote the start of input and the separation of RP with PL, to construct \emph{multi-granularity traffic sequence construction}, $x=\{[\text{CLS}]+[\text{RP}]+[\text{SEP}]+[\text{PL}]\}$. Then \emph{positional encoding} is added to $x$, and \emph{word embedding} is performed to convert the discrete sequence into a high-dimensional vector. 

\subsubsection{Traffic Contrastive Pre-training}
To overcome the traffic class imbalance and homogeneity problems, we design a \emph{positive-negative pairs sampling} of $x$ to conduct contrastive pre-training. The negative example of $x$ is set according to the following restrictive condition, \romannumeral1) sample from different classes as \emph{strong negative example} $y^{--}$; \romannumeral2) sample from the same class but with different communication information as \emph{weak negative example} $y^{-}$. The corresponding \emph{positive example} $y^{+}$ is obtained by enhancing $x$ with \emph{multi-head attention encoder} and \emph{projection layer}. The multi-head attention encoder consists of multi-head self-attention and feedforward neural networks, which encode contextual information and provide nonlinear variations, respectively. The projection layer provides nonlinear projection through fully connected layers using ReLu and then establishes a link between representations and a contrast loss function \emph{InfoNce}. Meanwhile, we focus more on strong negative examples by adjusting the loss weight of strong/weak in contrastive pre-training. 

\subsubsection{Semi-Supervised Fine-Tuning}
We propose to use unlabeled encrypted traffic for data re-balancing to mitigate class imbalance further. A fine-tuned $M_0$, initialized by pre-trained parameters of the multi-head attention encoder, predicts the unlabeled data to generate \emph{pseudo-labels}. The high-confidence parts are used along with the train dataset  to iteratively re-train $M_t$ until accuracy stabilizes. In addition, \emph{weighted sampling} and \emph{dynamic loss weighting} algorithms are used to alleviate the class imbalance of unlabeled data and focus more on training data than unlabeled data, ensuring that the iterated model \emph{does not} degrade the performance over the initial model. 

\section{Traffic Preprocessing}
We study existing methods \cite{wang2017malware, liu2019fs, zhang2019pccn, he2022payload} and conclude that most use only a single feature without sufficiently expressive, especially when modeling long-distance backgrounds. PASS designs a multi-granularity traffic sequence construction. The raw payload and packet length sequences are extracted by recoding to capture more information, then spliced in a specific way for word embedding.

\subsection{Multi-Granularity Traffic Sequence Construction}
The bidirectional flow $x$ is obtained according to the five-tuple from the original encrypted traffic. Depending on the existing work experience \cite{lotfollahi2020deep, lin2022bert} and sensitivity analysis, we extract the RP sequence of $m$=128 bytes and PL sequence of $n$=32 packets. 

\subsubsection{Raw Payload (RP) Sequence}
For each packet of $x$, we filter out bias information such as IP address, port number, Mac address, $etc.$, and take a total of 128 bytes (value range 0-255) \cite{rescorla2012datagram}. In order to obtain more context-sensitive information, we partition the RP sequence by two bytes, \emph{e.g.}, from \{1a, 2b, 03, 45, 62, aa, ...\} to \{1a2b, 0345, 62aa, ...\}. Similarly, the RP sequences of all flows are reconstructed and encoded into a coding dictionary (\emph{e.g.}, \{1a2b:0, 0345:1, 62aa:2\}) in the order of frequency of occurrence.

\subsubsection{Packet Length (PL) Sequence}
For each packet of $x$, we use “+” to indicate packets sent from the client to the server and “-” to indicate packets sent from the server to the client, \emph{e.g.},\{+328, -1074, -180, +328...\}, to represent the bi-directional nature of the flow. The length values of all flows in the dataset are also encoded and stored in the coding dictionary  (\emph{e.g.}, \{+328:3, -1074:4, -180:5\}) by frequency.

Four unique markers, \emph{i.e.}, CLS, SEP, PAD, and UNK, are added in the dictionary, which indicate the start flag of the input, the separator flag of the RP sequence and the PL sequence, the padding flag, and the unregistered word flag, respectively. Let the RP sequence after dictionary encoding as ${RP}=\left[r_0, \ldots, r_{m/2}\right]$, the PL sequence as $PL=\left[p_0, \ldots, p_n\right]$. The input of $x$ can be expressed as,
\begin{equation}
X=[CLS]+[r_0, \ldots, r_{m/2}]+[SEP]+[p_0, \ldots, p_n]
\end{equation}

\subsection{Word Embedding}
We perform word embedding of $x$ through the parameter matrix $W \in \mathbb{R}^{V \times d}$, which transforms the discrete input sequence $x \in \mathbb{R}^{(m/2+n+2) \times 1}$ into a high-dimensional vector $x \in \mathbb{R}^{(m/2+n+2) \times d}$, where $V$ denotes the size of $W$, and $d$ is the dimension. Additionally, the positional encoding information \cite{vaswani2017attention} is incorporated into the embedded vector to correct and enhance its contextual representation.

\section{Traffic Contrastive Pre-training}
Label bias is a common problem for class imbalance and homogeneous traffic classification tasks, which manifests in the form of blurred decision boundaries for homogeneous traffic and the tendency to divide minority classes into majority classes. PASS designs the positive-negative pairs sampling algorithm to expand the number of pre-training data, and alleviate label bias caused by the involvement of data distribution and label information by contrastive pre-training.

\subsection{Positive-Negative Pairs Sampling}
PASS constructs the positive and negative sample pairs to help learn minority classes and notice distinguishing overlapping features of homogeneous traffic, as shown in Alg.\ref{algorithm1}.
Specifically, the pre-trained data is a triple $\{x,x,y\}$, where $x$ is randomly selected and replicated once, and $y$ means $y^{--}$ or $y^{-}$. We consider different samples of $x$ that satisfy the following as $y$: 

\begin{itemize}
    \item Having different class labels as strong examples $y^{--}$, indicating that $x$ and $y$ are differentially labeled for the classification task and should be farther apart in the higher dimensional space. 
    \item Having the same label but different communication information, \emph{i.e.}, different TLS certificate or \emph{\{destination IP address, destination port\}} tuple, as weak negatives $y^{-}$, indicating that $x$ and $y$ show a low degree of homogeneity in communication patterns and should also be kept at a certain extent distance \cite{van2020flowprint, li2022foap}.
\end{itemize}

By using both types of negative examples, contrastive pre-training can focus on distinguishing not only between different app classes, but also similar communication patterns (\emph{i.e.}, homogeneous traffic) from different apps. Noteworthy, the communication information is not involved in classification tasks. We can focus more on the impact of strong negative examples by adjusting the loss weight of $y^{--}$ and $y^{-}$. See section \ref{LOSS} for details. 

Meanwhile, the positives of $x$ are obtained after data augmentation. $\{x, x\}$ will get two different representations after the multi-head attention encoder and projection layer due to the randomness of dropout, which can be regarded as the positives $y^{+}$ of each other. 

\SetKwFor{For}{for}{do}{endfor} 
\IncMargin{1em}
\begin{algorithm}
    \SetAlgoNoLine 
    \SetKwInOut{Input}{\textbf{Input}}\SetKwInOut{Output}{\textbf{Output}}\SetKwInOut{Require}{\textbf{Require}}
    \Input{Number of iterations $step$; Batch set size $bz$; $\mathcal{D}_{train}=\{(x_0, l_0, c_0), (x_1, l_1, c_n),$ $\ldots, (x_n, l_n, c_n)\}$, where $l_n$ as the label, $c_n$ as the communication information. 
        }
    \Output{Pre-training dataset $\mathcal{D}_{pre}$}

    \For {each $i \in[0, s t e p]$}{
          	$\operatorname{BZ}=[ \ ]$\\
            \For {each $j \in[0, b z]$}
            {
            	 $p=$ Random $([0, len(\mathcal{D})])$\\
            	 $state=1$\\
            	 \For {\_ in $range(500)$ and ${state==1}$}{
                	 $\triangleright$ \emph{500 times to deadlock prevention}\\
                	 $q=\operatorname{Random}([1, p-1] \cup[p+1, len(\mathcal{D})])$\\
                	 \If{($l_p \neq l_q$) or ($l_p==l_q$ and $c_p \neq c_q$)}
                	 {$\operatorname{BZ}.append\left(\{x_p, x_p, x_q\}\right)$\\$state=0$}
                	 % $\quad \triangleright$Positive sample class:${k}$, Negative sample class:${l}$\\
            	 }
           	}
           	$\mathcal{D}_{pre}.append(\operatorname{BZ})$\\
          }
    \caption{Positive-Negative Pairs Sampling\label{algorithm1}}
\end{algorithm}
\DecMargin{1em}

\subsection{Multi-Head Attention Encoder}
\begin{figure}[!htbp]
\centering
\includegraphics[scale=0.65]{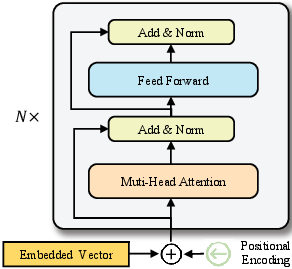}
\caption{Multi-head Attention Encoder Structure. Add \& Norm Means the Residual Network and Layer Normalization \cite{he2016deep, ba2016layer, devlin2018bert}.}
\label{fig2}
\end{figure}

The multi-head attention encoder consists of multi-head self-attention and feedforward neural networks, which can be stacked $N$=6 times to enhance the representation capability, as shown in Fig.\ref{fig2}. 
The multi-head self-attention network encodes contextual information, and the feedforward neural network provides nonlinear variation consisting of two linear layers and one layer of ReLu, which is connected by the Residual Networks \cite{he2016deep} and Layer Normalization \cite{ba2016layer}.

\subsection{Projection Layer and Contrast Loss Function}\label{LOSS}
The projection layer establishes a connection between the representation and the contrast loss by providing a nonlinear projection, through fully connected layers using the ReLu, to avoid the loss function missing some important features during pre-training \cite{miyato2018cgans}. Contrast pre-training aims to bring the positives closer and push the negatives further \cite{li2016deep}. We use the \emph{InfoNce} loss function to accomplish, calculated as follows, 
% 投影层通过使用 ReLu 的全连接层提供非线性投影，在表示和对比度损失之间建立联系，以避免损失函数在预训练期间丢失一些重要特征 \cite{miyato2018cgans, hsu2021hubert}。 对比预训练旨在使积极因素更接近并将消极因素推得更远 \cite{li2016deep}。 我们使用 \emph{InfoNce} 损失函数来完成上述目标，计算如下，
\begin{equation}
\begin{aligned}
L_{CPT}=L_{IN}\left(x, y_i^{+}\right) -\alpha * L_{IN}\left(x, y_i^{-/--}\right) \\
L_{IN}\left(x, y_i\right)=-\log \frac{\exp \left(\operatorname{sim}\left(x, y_i\right) / \tau\right)}{\sum_{k=1}^N \exp \left(\operatorname{sim}\left(x, y_k\right)\right)}
\end{aligned}
\end{equation}
where $\tau$ denotes the temperature parameter, and $\alpha$ denotes the weight of strong negatives. 
% 其中 $\tau$ 表示温度参数，$\alpha$ 表示强负样本的权重。

\section{Semi-supervised Fine-tuning}
Data re-balancing is the most direct and effective way to solve the class imbalance. Unlabeled and training flows come from the same scenario, where reliability and authenticity are sufficiently guaranteed. PASS proposes to use unlabeled encrypted flows for data re-balancing, effectively utilizing massive unlabeled traffic data while alleviating the dependence on labeled training data.
% 数据重平衡是解决类不平衡最直接有效的方法。无标签流和训练流来自同一个场景，可靠性和真实性得到充分保证。 PASS提出使用未标记的加密流进行数据再平衡，有效利用海量未标记流量数据，同时减轻对标记训练数据的依赖。

\subsection{Pseudo-Label Iteration}
We construct an initial classification model $M_0$, which adds a fully connected layer adapted to the classification task, initialized by the parameters of the pre-trained multi-head attention encoder and fine-tuned by the original train dataset. In each iteration, the current model $M_t$ predicts the unlabeled data to get the pseudo-labels, in which ones filtered by confidence threshold $thr$ and combined with the train dataset to re-train $M_{t+1}$. The iteration process is stopped when the effect increases insignificantly, \emph{e.g.}, $F1_{i+1}-F1_{i} \leq epsilon$, or reaches the upper number \emph{limit} of iterations. 

\subsection{Classification Model Re-training}
We propose weight sampling and dynamic loss weighting algorithms, to address the following problems when using pseudo-labeled data for re-training: 

\begin{itemize}
    \item Solve the pseudo-labeled data re-balancing. 
    \item Distinguish the contribution of pseudo-labeled and training data to the loss function. 
\end{itemize}

% \romannumeral1) solve the pseudo-labeled data re-balancing; \romannumeral2) distinguish the contribution of pseudo-labeled and training data to the loss function. 

\subsubsection{Weight Sampling}
The unlabeled data also have similar unbalanced proportions as the training data due to the consistency of data distribution in the same scenario. We sample more of the selected pseudo-label data for fewer original training data depending on the proportion of labels for each class, further alleviating the imbalance problem, 
\begin{equation}
s_i=\frac{1 / pro_i}{\sum_{i=1}^K 1 / pro_i}
\end{equation}
where $K$ denotes the number of classes in the train dataset, $pro_i$ is the proportion of data size per class to the total.

\subsubsection{Dynamic Loss Weighting}
The pseudo-labeled data will have some noise compared to the training data, \emph{i.e.}, the classification model will inevitably have some errors in the prediction even with the confidence threshold introduced. More attention must be focused on the training data during each iteration, which is reflected in the loss as different assigned weights \cite{dong2022leveraging}. 
Based on experience and some preliminary experimental results, the weight of the training data is $w_1=1$, and the pseudo-labeled data is $w_2=0.5$, where the learnable parameters will be updated according to gradient descent.

\section{Experiment}
In this section, we aim to answer the following questions:

\begin{itemize}
\item \textbf{RQ1:} Compared to existing ETC methods, can the proposed PASS achieve better classification performance, especially in combating class imbalance and homogeneous traffic?
% , especially in combating the label bias issues arising from class imbalance and homogeneous traffic?
\item \textbf{RQ2:} How does each part of PASS contribute to the performance? 
\item \textbf{RQ3:} How do sensitivity parameters, such as the number of pre-training samples, affect the performance?
\item \textbf{RQ4:} For existing ETC methods, can PASS provide some guidance to improve their performance?
\end{itemize}

\subsection{Experiment Setting}\label{es}
\subsubsection{Dataset Description}
We utilize four publicly available datasets in our experiments:

\begin{itemize}
\item \textbf{Cross-Platform215 (C-P215)} \cite{ren2019international}, is a dataset of traffic for popular apps on the Android platform in China, the United States, and India for 215 classes.
\item \textbf{ISCX-17} \cite{draper2016characterization}, includes 7 types of traffic for VPN and non-VPN communications, combined by apps, resulting in 17 different classes. 

\item \textbf{CST-TLS1.3 (C-TLS1.3)} \cite{lin2022bert}, is a dataset of services, websites, apps focused on the TLS1.3 protocol, with 120 classes. 

\item \textbf{CICInvesAndMal2019 (CIC2019)} \cite{taheri2019extensible}, is a dataset of 5 families of malicious traffic and 1 normal traffic.

\end{itemize}

\begin{table}[!htbp]
\captionsetup{font={footnotesize}}
\renewcommand\arraystretch{0.8}
\scriptsize
\caption{The Statistical Information of the Datasets, and the Maximum Class Imbalance Ratio t (MCIR) of Datase.}
\centering
\resizebox{\linewidth}{!}{
\begin{tabular}{c|ccc}
\toprule
\textbf{\textcolor{white}{--------}Dataset\textcolor{white}{--------}}  & \textbf{Flow} & \textbf{Label} & \textbf{MCIR}
\\ \cmidrule{1-4}
{C-P215} & 39,641 & 215 & 1,634:3
% (136.17) 
\\ \cmidrule{1-4}
{ISCX-17} & 14,728 & 17 & 4,315:42
% (102.73)
\\ \cmidrule{1-4}
{C-TLS1.3} & 46,372 & 120  & 500:16
% (31.25)  
\\ \cmidrule{1-4}
{\textcolor{white}{----}CIC2019\textcolor{white}{----}} 
& \textcolor{white}{----}13,972   \textcolor{white}{----}
& \textcolor{white}{----}6        \textcolor{white}{----}
& \textcolor{white}{----}4,998:653\textcolor{white}{----}
% (7.65)
\\ \bottomrule
\end{tabular}}
\label{Tab1}
\end{table}
%%%%%%%%%%%%%%%%%%%%%%%%%%%%%%%%%%%%%%%%%%%%%%%%%%%%%%%%%%%%%%%%%%%%%%%%
\begin{figure} [!htbp]
	\centering
	\captionsetup{font={footnotesize}}
	\subfloat[]{\includegraphics[scale=0.2]{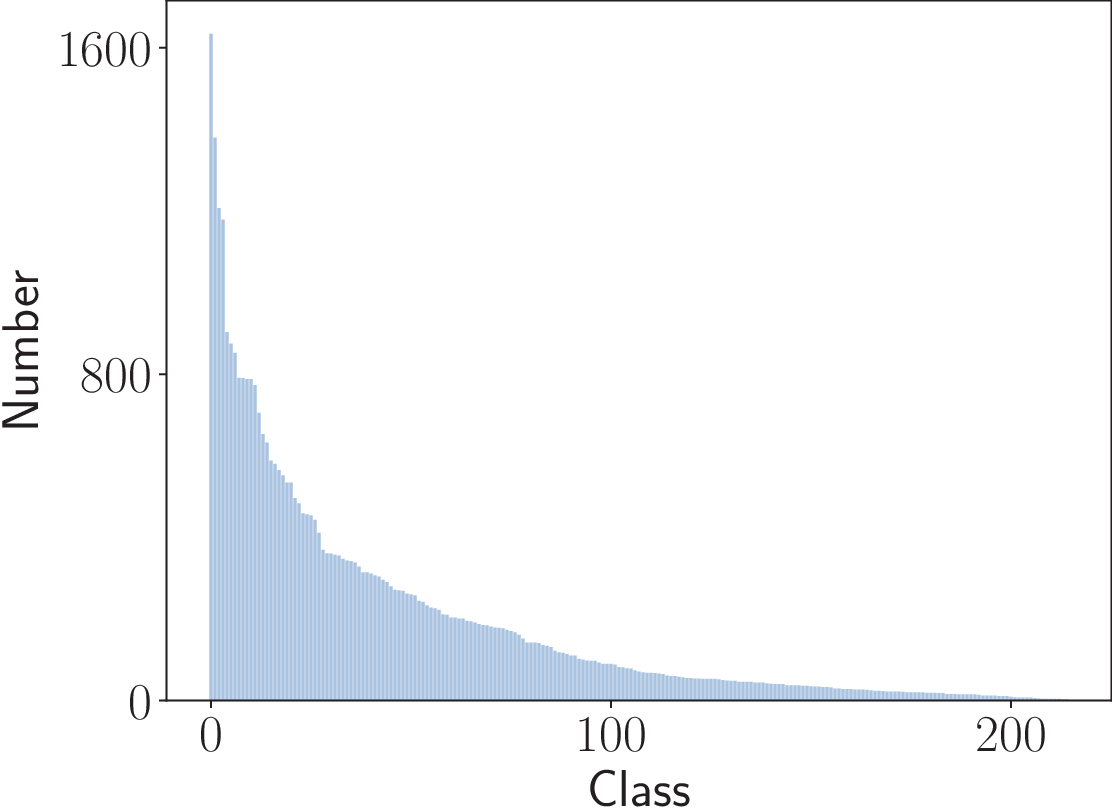}}
		\hspace{0.30cm}
	\subfloat[]{\includegraphics[scale=0.2]{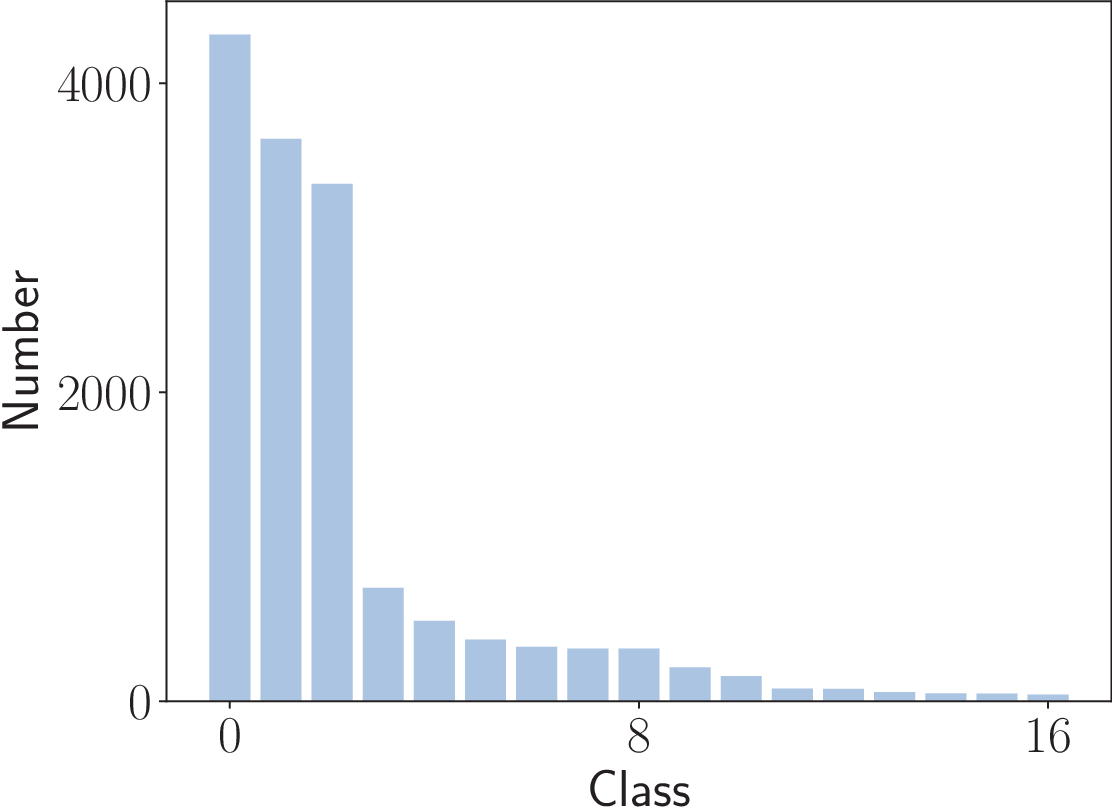}}
		\hspace{0.30cm}
	\subfloat[]{\includegraphics[scale=0.2]{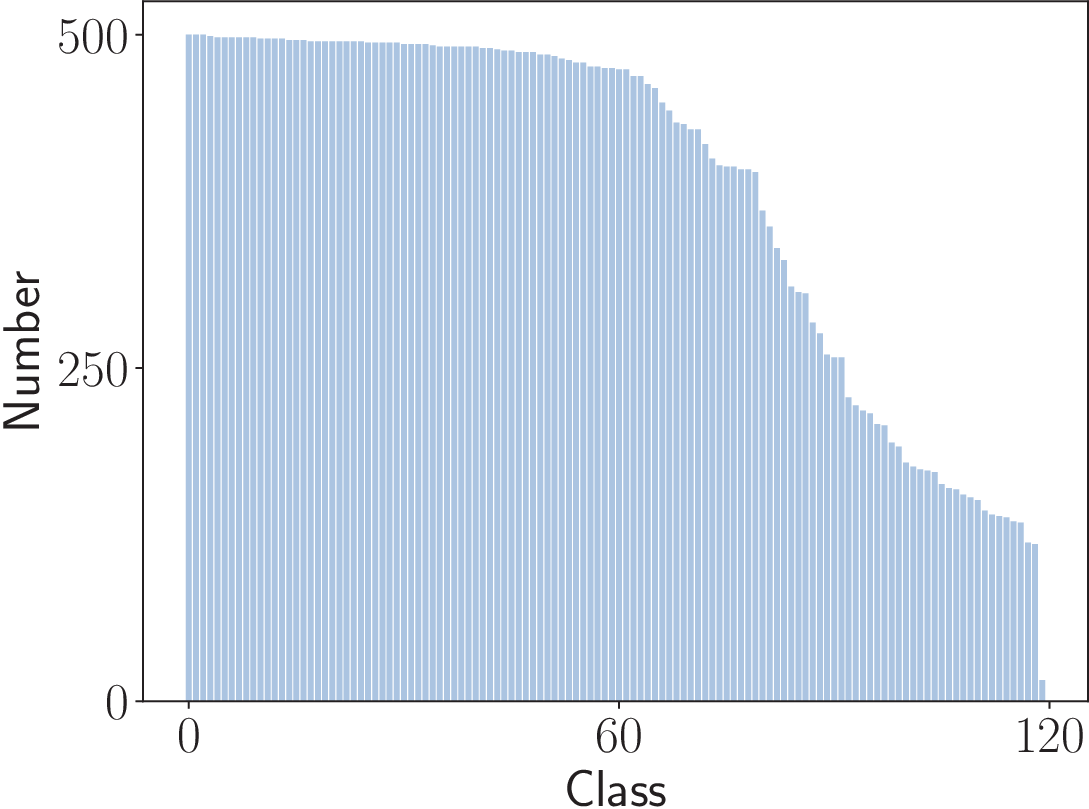}}
		\hspace{0.30cm}
	\subfloat[]{\includegraphics[scale=0.2]{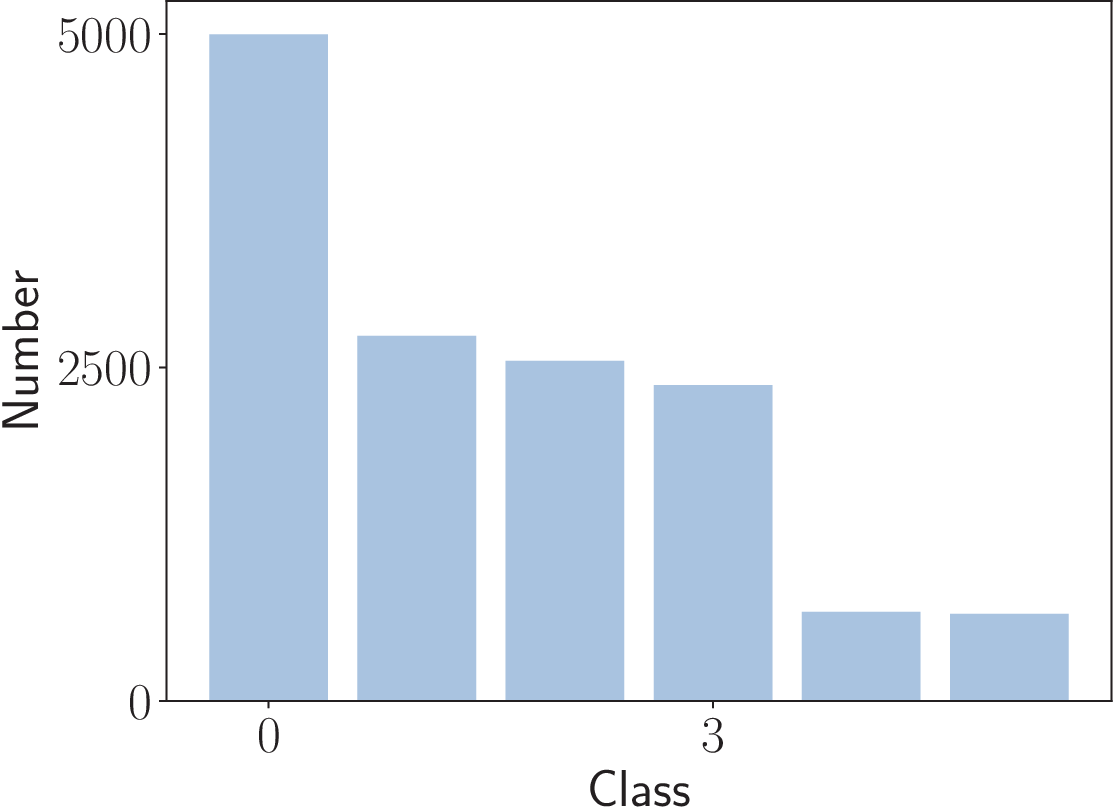}}
	\caption{Number of Each Class in the Dataset. 
	(a) C-P215, (b) ISCX-17, (c) C-TLS1.3, and (d) CIC2019.}
	\label{fig3} 
\end{figure}
%%%%%%%%%%%%%%%%%%%%%%%%%%%%%%%%%%%%%%%%%%%%%%%%%%%%%%%%%%%%%%%%%%%%%%%%

%%%%%%%%%%%%%%%%%%%%%%%%%%%%%%%%%%%%%%%%%%%%%%%%%%%%%%%%%%%%%%%%%%%%%%%%
\begin{table}[!htp]
\captionsetup{font={footnotesize}}
\renewcommand\arraystretch{0.75}
\scriptsize
\caption{The Number of Apps with Homogeneous Domains across the US, China, and India Markets.}
\centering
\resizebox{\linewidth}{!}{
\begin{tabular}{c|ccc|cc}
\toprule
\textbf{Domain}  & \textbf{US}	&  \textbf{China}		& \textbf{India} & \textbf{Sum}		& \textbf{Percentage}
\\ \cmidrule{1-6}
{google.com}            & 78 & 46 & 79 & 203 & 94.42\% \\ \cmidrule{1-6}
{crashlytics.com}       & 49 & 10 & 58 & 117 & 54.42\%\\ \cmidrule{1-6}
{googleapis.com}        & 16 & 71 & 0  & 87  & 40.47\%\\ \cmidrule{1-6}
{googleadservices.com}  & 18 & 14 & 34 & 66  & 30.70\%\\ \cmidrule{1-6}
{qq.com}                & 1  & 48 & 1  & 50  & 23.26\%\\ \cmidrule{1-6}
{cloudfront.net}        & 14 & 0  & 31 & 45  & 20.93\%\\ \cmidrule{1-6}
{appsflyer.com}         & 25 & 4  & 16 & 45  & 20.93\%\\ \cmidrule{1-6}
{umeng.com}             & 5  & 38 & 2  & 45  & 20.93\%\\ \cmidrule{1-6}
{doubleclick.net}       & 20 & 6  & 17 & 43  & 20.00\%\\ \cmidrule{1-6}
{baidu.com}             & 0  & 41 & 0  & 41  & 19.07\%
\\ \cmidrule{1-6}
{amazonaws.com}         & 13 & 0  & 20 & 33  & 15.35\%\\ \cmidrule{1-6}
{taobao.com}            & 0  & 29 & 1  & 30  & 13.95\%\\ \cmidrule{1-6}
{amap.com}              & 0  & 26 & 0  & 26  & 12.09\%\\ \cmidrule{1-6}
{qlogo.cn}              & 0  & 24 & 0  & 24  & 11.16\%\\ \cmidrule{1-6}
{p3-group.com}          & 0  & 0  & 22 & 22  & 10.23\%
\\ \cmidrule{1-6}
{scorecardresearch.com} & 10 & 1  & 11 & 22  & 10.23\%
\\ \cmidrule{1-6}
branch.io             & 11 & 0  & 10 & 21  & 9.77\%\\ 
\cmidrule{1-6}
moatads.com           & 12 & 1  & 4  & 17  & 7.91\%\\ 
\cmidrule{1-6}
irs01.com             & 0  & 17 & 0  & 17  & 7.91\%\\ 
\cmidrule{1-6}
igexin.com            & 0  & 17 & 0  & 17  & 7.91\%
\\ \bottomrule
\end{tabular}}
\label{Tab2}
\end{table}
%%%%%%%%%%%%%%%%%%%%%%%%%%%%%%%%%%%%%%%%%%%%%%%%%%%%%%%%%%%%%%%%%%%%%%%%

As shown in Table.\ref{Tab1}, the traffic distribution of all four datasets shows an apparent long-tail distribution with class imbalance ratios between 8:1 to 545:1, where the specific value of each class is shown in Fig.\ref{fig3}.
Meanwhile, each dataset has different degrees of traffic homogeneity phenomenon. We use the app number of the same app family and the percentage of the same domain appearing in different apps to measure the traffic homogeneity in a dataset. 
For instance, on the C-P215 dataset, the Tencent family consists of at least 9 representative apps such as \emph{com.tencent.news} and \emph{com.qq.reader}, while the Amazon family has at least 4 apps such as \emph{com.amazon.kindle} and \emph{com.amazon.tahoe}. 
The dataset also contains a significant amount of traffic from the same cloud provider domains, such as \emph{google.com} and \emph{umeng.com}, as shown in Table \ref{Tab2}, which displays the top 20 domains that appear in multiple apps.

\subsubsection{Implementation Details and Evaluation Metrics}
All experiments are implemented using Pytorch and performed on the Telsa V100 GPUs. The multi-head attention encoder contains 8 heads, the vector dimension of $q_i$, $k_i$, $v_i$=64, and the number of neurons in the feedforward neural network is 1,024. We choose the BertAdam optimizer with a learning rate of 5e-5 and a warmup of 0.03. Hyperparameters are selected by performing a grid search calibration procedure with cross-validation,
the weight of strong negatives $\alpha$=2, the upper number of pseudo-label iterations \emph{limit}=5, and the $thr$=0.95.

We evaluate and compare the performance by four typical metrics, including Accuracy (AC), Precision (PR), Recall (RC), and F1 score. Macro Average \cite{zheng2020learning} is used to avoid biased results due to imbalance between multiple classes of data by calculating the mean value of AC, PR, RC, and F1 of each class. Each dataset is divided into the train, validation, and test dataset, with a ratio of 8:1:1. We sample each train dataset with a sample size of about 500,000 to obtain the pre-training dataset, and perform evaluations on the test dataset to report metrics.

% 所有的实验都用Pytorch实现，并在Telsa V100 GPU上进行。多头注意力编码器包含8个头，$q_i$、$k_i$、$v_i$的向量维度为64，前馈神经网络的神经元数量为1024。我们选择BertAdam优化器，学习率为5e-5，预热为0.03。超参数是通过执行网格搜索校准程序和交叉验证来选择的。伪标签迭代次数的上限是5，本文中$thr$被选为0.95。我们通过四个典型的指标来评估和比较性能，包括准确率（AC）、精确度（PR）、召回率（RC）和 F1得分。Macro Average\cite{liu2017efficient}被用来避免由于多类数据之间的不平衡而产生的偏颇结果，计算每个类的AC、PR、RC和F1分数的平均值。每个数据集被分为训练、验证和测试数据集，比例为8：1：1。同时，我们对每个训练数据集进行抽样，样本量约为500,000，以获得预训练数据集，并对测试数据集进行评估，以报告指标。 

\begin{table*}[htbp]
\captionsetup{font={normalsize}}
\caption{Comparison Results on C-P215, ISCX-17, C-TLS1.3, CIC2019.}
\label{Tab3}
\renewcommand\arraystretch{1.1}
\resizebox{\linewidth}{!}{
\begin{tabular}{c|cccc|cccc|cccc|cccc}
\toprule
\textbf{Dataset}
& \multicolumn{4}{|c}{\textbf{C-P215}} 
& \multicolumn{4}{|c}{\textbf{ISCX-17}} 
& \multicolumn{4}{|c}{\textbf{C-TLS1.3}} 
& \multicolumn{4}{|c}{\textbf{CIC2019}}
\\
\cmidrule[0.8pt](r){1-17}
\textbf{Method}
&{AC}	&{PC}		&{RC}		&{F1}
&{AC}	&{PC}		&{RC}		&{F1}
&{AC}	&{PC}		&{RC}		&{F1}
&{AC}	&{PC}		&{RC}		&{F1}
\\
\cmidrule[0.8pt](r){1-17}
{AppScanner}
& 38.63 & 25.23 & 25.94 & 24.40 

& 62.66 & 48.64 & 51.98 & 49.35 

& 66.62 & 62.46 & 63.27 & 62.01 

& 57.32 & 54.36 & 53.87 & 53.52
\\
\cmidrule(r){1-1} \cmidrule(r){2-5} \cmidrule(r){6-9} \cmidrule(r){10-13} \cmidrule(r){14-17}
{2D-CNN}
& 92.43 & 85.04 & 81.14 & 81.63 

& 67.00 & 65.18 & 63.38 & 63.96 

& 89.79 & 87.14 & 86.42 & 86.12 

& 84.15 & 82.45 & 82.71 & 82.56
\\
{FS-Net}
& 48.46 & 35.44 & 33.65 & 33.43 

& 66.47 & 48.19 & 48.48 & 47.37 

& 86.39 & 84.04 & 83.49 & 83.22 

& 89.17 & 87.08 & 87.12 & 86.67
\\
{BiLSTM}
& 87.42 & 84.26 & 82.45 & 82.54 

& 63.56 & 62.91 & 62.37 & 62.83 

& 89.41 & 89.57 & 87.41 & 87.74 

& 85.91 & 84.50 & 84.22 & 84.18
\\
{PCNN}
& 98.76 & 93.87 & 94.01 & 93.57 

& 77.02 & 74.38 & 75.57 & 73.97 

& 93.75 & 93.28 & 92.79 & 93.10 

& 91.39 & 88.45 & 90.73 & 89.57
\\
\cmidrule(r){1-1} \cmidrule(r){2-5} \cmidrule(r){6-9} \cmidrule(r){10-13} \cmidrule(r){14-17}
{PERT}
& 97.72 & 86.28 & 85.91 & 85.50 

& 82.29 & 70.92 & 71.73 & 69.92 

& 89.15 & 88.46 & 87.19 & 87.41 

& 90.37 & {89.49} & 88.92 & 88.26
\\
{ET-Bert}
& 98.65 & 93.24 & 92.66 & 92.46

& 85.19 & 75.08 & 72.94 & 73.06

& \textbf{95.10} & 94.60 & \textbf{94.19} & \textbf{94.26}

& 91.89 & 89.23 & 91.76 & 90.28
\\
\cmidrule[0.8pt](r){1-17}
\textbf{PASS}
& \textbf{98.98} & \textbf{95.17} & \textbf{95.23} & \textbf{94.88} 

& \textbf{87.55} & \textbf{83.11} & \textbf{84.11} & \textbf{83.09} 

& 94.85 & \textbf{94.78} & {94.01} & \textbf{94.26}

& \textbf{93.82} & \textbf{91.66} & \textbf{95.28} & \textbf{93.19}
\\
\bottomrule
\end{tabular}}
\end{table*}

\subsection{Comparison with State-of-the-Art Methods (RQ1)}\label{Sort}
We compare PASS with 7 state-of-the-art ETC methods, including,
\begin{itemize}
\item \textbf{Statistical-feature-based}: AppScanner \cite{taylor2016appscanner}.
\item \textbf{Sequence-feature-based}: 2D-CNN \cite{wang2017malware}, Fs-Net \cite{liu2019fs}, BiLSTM \cite{huang2015bidirectional}, and PCNN \cite{zhang2019pccn}. 
\item \textbf{Pre-training-based}: PERT \cite{he2022payload}, ET-BERT \cite{lin2022bert}. 
\end{itemize}

As can be seen the comparison results from Table.\ref{Tab3}, PASS significantly outperforms all statistical feature-based and sequence feature-based methods. Compared with the other two pre-training-based methods, PASS achieves the most notable improvement on the C-P215 dataset, which has the most significant class imbalance issue, with an F1 increase of 2.42\% and 9.38\%, respectively. Additionally, PASS performs competitively on the C-TLS1.3 dataset with less class imbalance. PASS only samples 500,000 encrypted traffic flows for contrastive pre-training on each dataset, but ET-BERT collects 30GB of external unlabeled pre-training data. Nevertheless, in most cases, PASS outperforms ET-BERT mainly for the following reasons,
\begin{itemize}
    \item Contrastive pre-training optimization target adopted by PASS is better suited for addressing the class imbalance and traffic homogeneity issues in ETC scenarios compared to general masked language model pre-training task used by ET-BERT.
    \item PASS performs data re-balancing, which helps adjust the distribution of unlabeled data and improves the learning of the traffic characteristics of minority apps.
\end{itemize}

%%%%%%%%%%%%%%%%%%%%%%%%%%%%%%%%%%%%%%%%%%%%%%%%%%%%%%%%%%%%%%%%%%%%%%%%
\begin{figure*} [!]
	\centering
	\captionsetup{font={footnotesize}}
	\subfloat[]{\includegraphics[height=29mm,width=36mm]{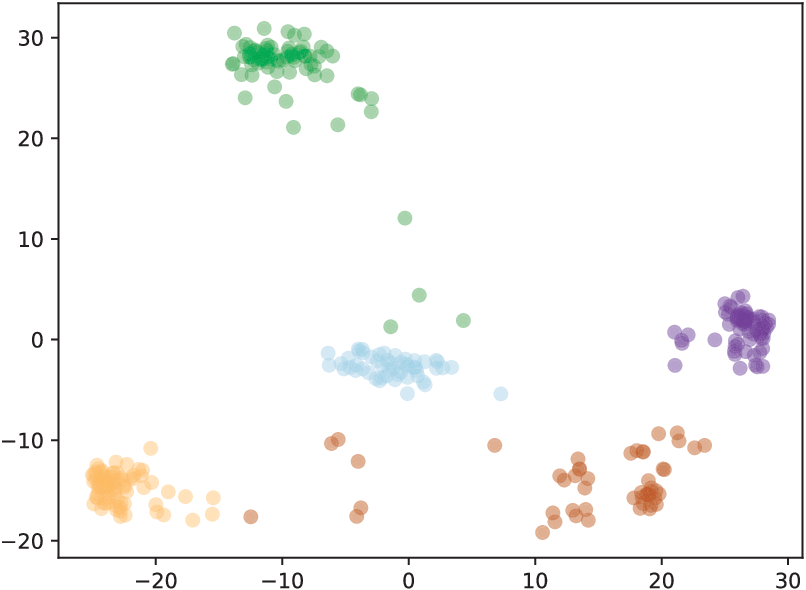}}
		%\hspace{0.2cm}
	\subfloat[]{\includegraphics[height=29mm,width=36mm]{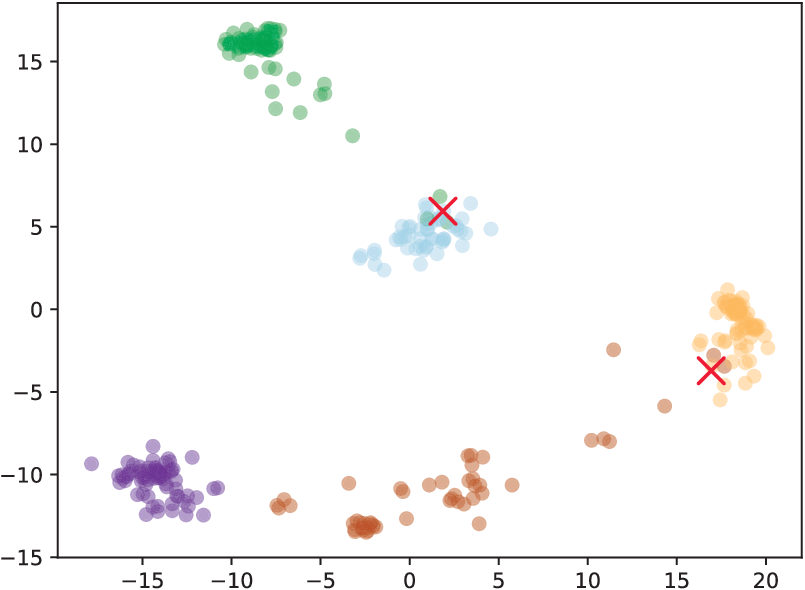}}
		%\hspace{0.2cm}
	\subfloat[]{\includegraphics[height=29mm,width=36mm]{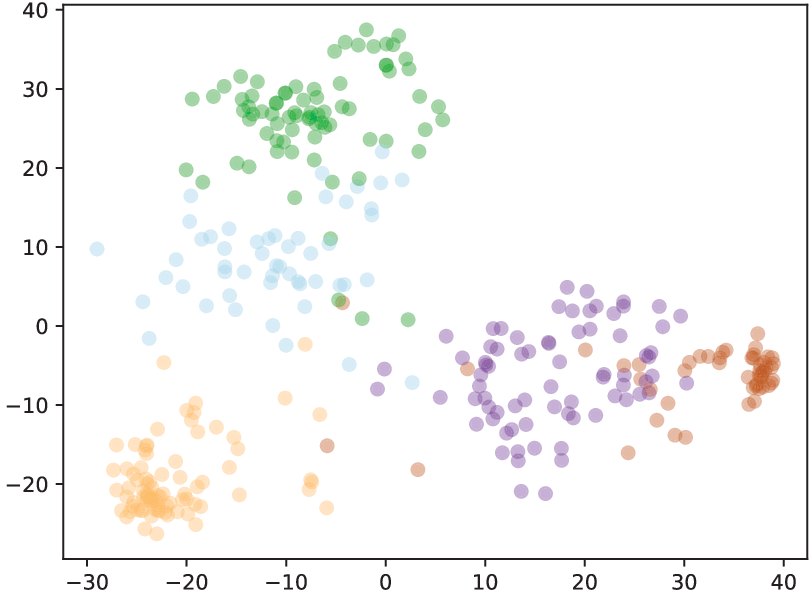}}
		%\hspace{0.2cm}
	\subfloat[]{\includegraphics[height=29mm,width=36mm]{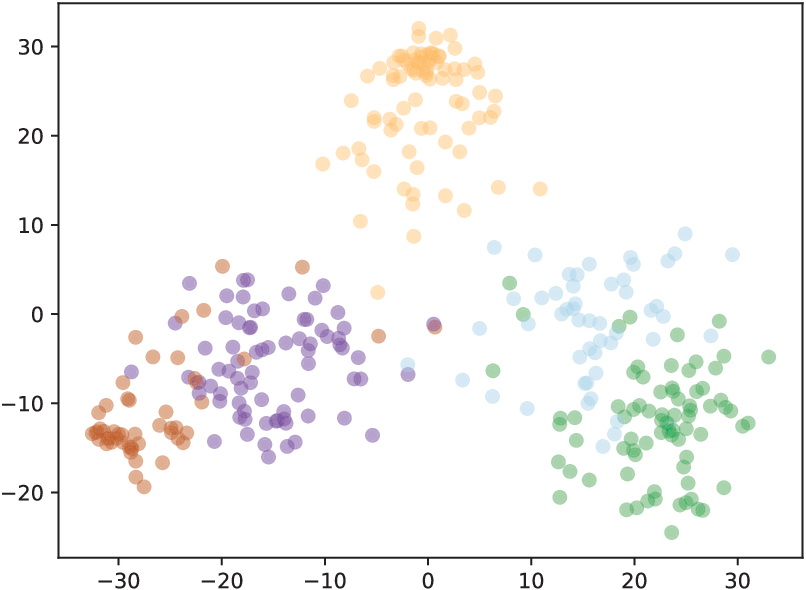}}
		%\hspace{0.2cm}
	\subfloat[]{\includegraphics[height=29mm,width=36mm]{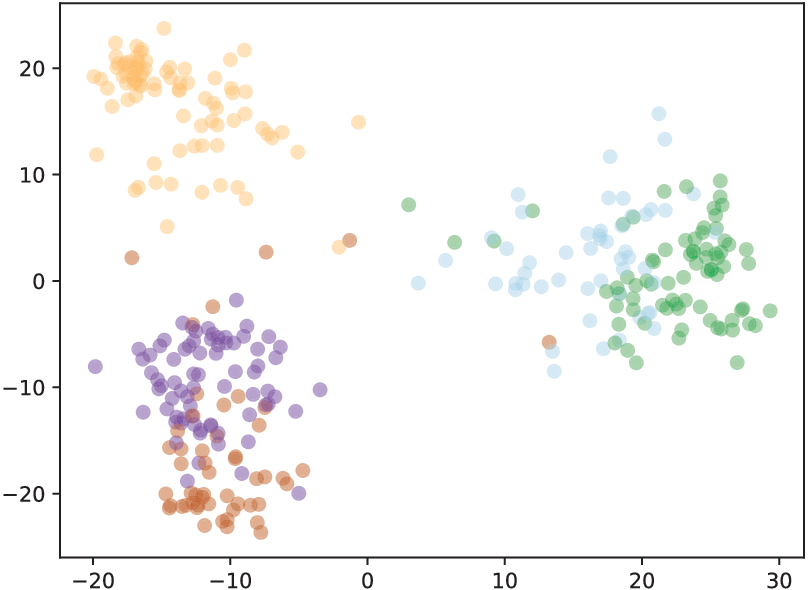}}
	\caption{Visualizations of Classification Vectors. (a) PASS, (b) \emph{w/o} PLI,	(c) \emph{w/o} PLI+CPT, (d) \emph{w/o} PLI+CPT+PL, and (e) \emph{w/o} PLI+CPT+RP.}
	\label{fig4} 
\end{figure*}
%%%%%%%%%%%%%%%%%%%%%%%%%%%%%%%%%%%%%%%%%%%%%%%%%%%%%%%%%%%%%%%%%%%%%%%%

\subsection{Ablation Study (RQ2)}\label{AS}
\subsubsection{Ablation Study of Key Components}
We perform ablation tests to verify the contribution of key components. Specifically, we unloaded the Pseudo-Label Iteration (PLI), Contrastive Pre-Training (CPT), and Raw payload (RP)/Packet Length (PL) sequence feature in turn, the results as shown in Table.\ref{Tab4}. The following three phenomena can be observed:
% 我们执行消融测试以验证关键组件的贡献。具体来说，我们依次卸载伪标签迭代（PLI）、对比预训练（CPT）和原始载荷（RP）/数据包长度（PL）序列特征，实验结果如表所示。\ref{Tab4}\footnote{我们将数据集名称缩写为更好的表格排版。}。可以观察到以下三种现象：
\begin{itemize}
    \item Eliminating the PLI module leads to significant performance decreases of PASS on all datasets, with an average drop of 3.27\% in the F1. The decline is most evident on the ISCX-17 dataset, where the F1 dropped by 8.38\%.
    \item The negative impact caused by removing CPT is relatively consistent, with a max-decrease of 1.72\% and a mini-decrease of 0.99\%; in other words, increasing pre-training can steadily improve the ETC model performance. 
    \item The adverse effects of ablating features are unpredictable, which proves that different ones contribute differently to various datasets and further shows that multi-granularity features will be more robust than single-granularity. 
\end{itemize}

\begin{table}[htp]
\centering
\scriptsize
\captionsetup{font={footnotesize}}
\caption{Ablation Study of Key Components on F1. \textcolor{red}{Red} indicates the difference from the previous column. \emph{w/o} means without. Noteworthy, \emph{w/o} PLI+CPT+RP and \emph{w/o} PLI+CPT+PL are compared with \emph{w/o} PLI+CPT.
}
\label{Tab4}
\renewcommand\arraystretch{0.9}
\resizebox{\linewidth}{!}{
\begin{tabu}{c|cccc}
\toprule
\textbf{Method}    
& {\textbf{C-P215}} 

& {\textbf{ISCX-17}} 

& {\textbf{C-TLS1.3}} 

& {\textbf{CIC2019}}
\\ \cmidrule(r){1-5} 
PASS
& \textbf{94.88}

& \textbf{83.09}

& \textbf{94.26}

& \textbf{93.19}
\\ \cmidrule(r){1-5} 
\emph{w/o} PLI
& 93.73/\textcolor{red}{1.15$\downarrow$}

& 74.71/\textcolor{red}{8.38$\downarrow$}

& 93.62/\textcolor{red}{0.64$\downarrow$}

& 90.30/\textcolor{red}{2.89$\downarrow$}
\\ \cmidrule(r){1-1} \cmidrule(r){2-5}
\emph{w/o} PLI+CPT
& 92.01/\textcolor{red}{1.72$\downarrow$}

& 73.45/\textcolor{red}{1.26$\downarrow$}

& 92.40/\textcolor{red}{1.22$\downarrow$}

& 89.31/\textcolor{red}{0.99$\downarrow$}
\\ \cmidrule(r){1-1} \cmidrule(r){2-5}
\emph{w/o} PLI+CPT+PL
& 91.16/\textcolor{red}{0.85$\downarrow$}

& 70.25/\textcolor{red}{3.20$\downarrow$}

& 90.43/\textcolor{red}{1.97$\downarrow$}

& 89.12/\textcolor{red}{0.19$\downarrow$}
\\ \cmidrule(r){1-1} \cmidrule(r){2-5}

\emph{w/o} PLI+CPT+RP
& 89.73/\textcolor{red}{2.28$\downarrow$}

& 65.96/\textcolor{red}{7.49$\downarrow$}

& 89.27/\textcolor{red}{3.13$\downarrow$}

& 89.27/\textcolor{red}{0.04$\downarrow$}

\\ \bottomrule
\end{tabu}}
\end{table}

\begin{table}[htp]
\centering
\scriptsize
\captionsetup{font={footnotesize}}
\caption{Comparison of Generic Sampling Approaches with \emph{w/o} PLI on F1. \textcolor{orange}{Orange} indicates the difference from the highest value. }

\label{Tab5}
\renewcommand\arraystretch{0.9}
\resizebox{\linewidth}{!}{
\begin{tabu}{c|cccc}
\toprule
\textbf{Method}    
& {\textbf{C-P215}} 
& {\textbf{ISCX-17}} 
& {\textbf{C-TLS1.3}} 
& {\textbf{CIC2019}}

\\\midrule
% \emph{w/o} PLI+CPT\textcolor{white}{------------}
% & 92.01
% & 73.45
% & 92.40
% & 89.31

% \\ \cmidrule(r){1-5}

\textcolor{white}{++PL}\emph{w/o} PLI\textcolor{white}{CPT}
& \textbf{93.73}
& {74.71}/\textcolor{orange}{1.82$\downarrow$}
& \textbf{93.62}
& \textbf{90.30}
\\ \cmidrule(r){1-5}

ROS
& 93.43/\textcolor{orange}{0.30$\downarrow$}
& 73.92/\textcolor{orange}{2.61$\downarrow$}
& 92.57/\textcolor{orange}{1.05$\downarrow$}
& 86.35/\textcolor{orange}{3.95$\downarrow$}

\\ \cmidrule(r){1-1} \cmidrule(r){2-5}
RUS
& 92.81/\textcolor{orange}{0.92$\downarrow$}
& 68.15/\textcolor{orange}{6.56$\downarrow$}
& 85.78/\textcolor{orange}{8.38$\downarrow$}
& 84.91/\textcolor{orange}{5.39$\downarrow$}

\\ \cmidrule(r){1-1} \cmidrule(r){2-5}
SMOTE
& 92.15/\textcolor{orange}{1.58$\downarrow$}
& \textbf{76.53}
& 92.97/\textcolor{orange}{0.65$\downarrow$}
& 87.80/\textcolor{orange}{2.50$\downarrow$}

\\ \cmidrule(r){1-1} \cmidrule(r){2-5}
FocalLoss
& 93.70/\textcolor{orange}{0.03$\downarrow$}
& 74.77/\textcolor{orange}{1.76$\downarrow$}
& 92.57/\textcolor{orange}{1.05$\downarrow$}
& 87.49/\textcolor{orange}{2.81$\downarrow$}

\\ \bottomrule
\end{tabu}}
\end{table}

\subsubsection{Ablation Study of Class Imbalance}
PASS achieves sampling by constructing positive-negative pairs for pre-training. In order to verify the superiority of our sampling approach, we equip the following four sampling approaches to \emph{w/o} PLI+CPT for comparison: 

\begin{itemize}
    \item Random Over Sampling (ROS): random replication of minority class samples based on the re-sampling \cite{japkowicz2000learning}.
    \item Random Under Sampling (RUS): random removal of majority class samples based on the down-sampling \cite{drummond2003c4}.
    \item SMOTE: generate new samples by linearly combining the nearest neighbor samples of minority classes \cite{chawla2002smote}. 
    \item FocalLoss: optimize hard-to-classify and easy-to-classify samples due to class imbalance by assigning different weights to the loss function based on cost sensitivity \cite{lin2017focal}. 
\end{itemize}

Combining the results in Table.\ref{Tab4} and Table.\ref{Tab5}, it shows that compared with \emph{w/o} PLI+CPT, SMOTE increases by 0.57\% on average, and FocalLoss increases by 0.34\%, but  RUS decreases on multiple datasets, suggesting that downsampling is only suitable for numerous sizes of minorities. The \emph{w/o} PLI outperforms the above sampling schemes except for ISCX-17, because the small size of ISCX-17 causes some overfitting, while enlightens that the class imbalance problem of small size can be addressed by synthesizing new samples or adjusting the loss function.

\subsubsection{Ablation Study of Traffic Homogeneity}
We select 5 apps from Tencent family,
\{\emph{com.tencent.mobileqq}, \emph{com.qq.reader}, \emph{com.tencent.news}, \emph{com.tencent.qqmusic}, \emph{com.tencent}\emph{.radio}\}, 
and top 5 homogeneous domains in Table.\ref{Tab2}, from the C-P215, in order to construct the homogeneous app family and homogeneous domain test sets to perform app label classification, respectively, further investigating the contribution of each component to classify homogeneous traffic. 
For better visualization, we use t-SNE to plot the classification layer vectors of the Tencent family traffic, as shown in Fig.\ref{fig4}, which demonstrates that PASS generates more separable representations. 
We present the app label classification results for homogeneous domains in Fig.\ref{fig5}. The RP sequence feature is more beneficial for the classification task of homogeneous domain traffic, as the AC of \emph{google.com}, \emph{crashlytics.com}, and \emph{googleapis.com} significantly decreases when it is removed.

\begin{figure}[!htbp]
\centering
\includegraphics[scale=0.6]{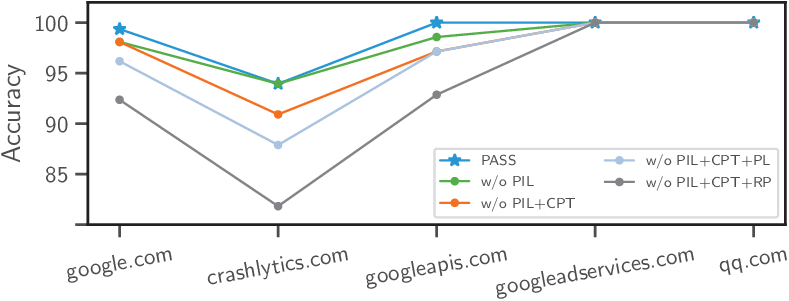}
\caption{App Label Classification Accuracy of Homogeneous Domain Traffic.}
\label{fig5}
\end{figure}
% PASS              0.9906, 0.9870, 0.9795, 0.9808
% w/o PIL           0.9874, 0.9776, 0.9754, 0.9744
% w/o PIL+CPT       0.9749, 0.9572, 0.9532, 0.9510
% w/o PIL+CPT+PL    0.9687, 0.9416, 0.9417, 0.9380
% w/o PIL+CPT+RP    0.9278, 0.8596, 0.8773, 0.8628

\subsection{Sensitivity Analysis (RQ3)}
We show the sensitivity analysis result to investigate the effect of the number of pre-training samples and the combination of RP/PL sequence lengths on the widely compared C-P215. The number of pre-training is sequentially selected from [30,000, 40,000, ..., 80,000], and the average F1 is after fine-tuning. From the Fig.\ref{fig5} (a), we can observe the following:
\begin{itemize}
    \item The number of pre-trained samples is linearly related to training time. 
    \item Pre-train 500,000 samples take about 1 hour, which scales to 600,000 for optimal F1, with a corresponding significant increase in training time.
    \item After 600,000, the increasing trend is not obvious, mainly due to the lack of diversity in the pre-training samples obtained by re-sampling. 
\end{itemize}

Meanwhile, the combination of RP/PL sequence lengths [$m$, $n$] is sequentially selected from  = \{[8, 4], [16, 8], [32, 16], [64, 32], [128, 64]\}. We find an overlapping phenomenon from Fig.\ref{fig5} (b), the value of F1 is getting higher, but the training time is also increasing. 
Although the optimal F1 is 96.27, we recommend setting [$m/2$, $n$] to [64, 32], considering the training efficiency and the difficulty of storing sufficient packets for a five-tuple flow in the real-world network environment.

\begin{figure}[htbp]
\centering
\subfloat[]{\includegraphics[scale=0.33]{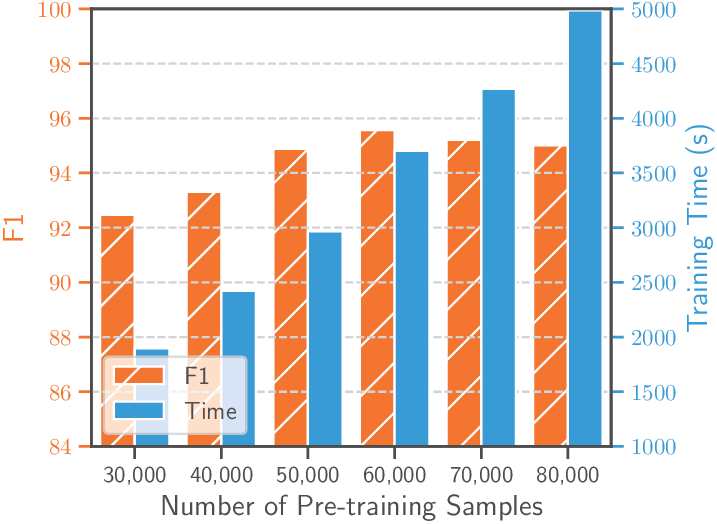}%
\label{fig7_first_case}}
\hspace{0.01cm}
\subfloat[]{\includegraphics[scale=0.33]{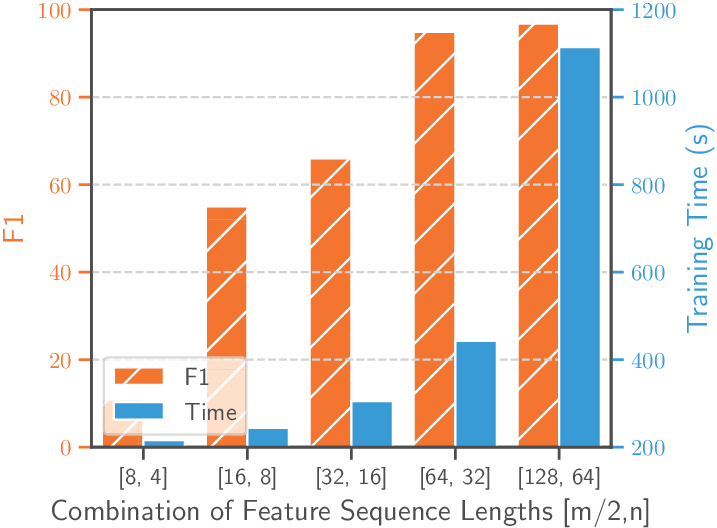}%
\label{fig7_second_case}}
\caption{Sensitivity Analysis Results. (a) Respective Relationships between F1, the Number of Pre-training Samples, and the Training Time; (b) Respective Relationships between F1, the Combination of Sequence Feature Lengths, and the Training Time. }
\label{fig5}
\end{figure}

\subsection{Scalability Analysis (RQ4)}
To investigate the scalability of the PLI and CPT components, we apply them to two classic ETC models: 2D-CNN and BiLSTM, both of which have poor performance on the ISCX-17 and CIC2019 datasets. Specifically, we evaluate the degree of improvement in F1 when adding the PLI and CPT components to these models. The experimental results are shown in Table \ref{Tab6}, where we observe an average improvement of 4.59\% based on the 2D-CNN model and 4.29\% based on the BiLSTM model. These results demonstrate that the combination of contrastive pre-training and pseudo-label iteration can effectively enhance the performance of existing ETC models, showcasing the scalability of PASS and its potential to guide other ETC-related work.

\begin{table}[hpt]
\captionsetup{font={footnotesize}}
\caption{F1 of 2D-CNN and BiLSTM F1, equipped with PLI and CPT, on ISCX-17 and CIC2019 datasets.}
\label{Tab6}
\renewcommand\arraystretch{1}
\resizebox{\linewidth}{!}{
\begin{tabular}{c|ccc|ccc}
\toprule
\textbf{Dataset}
& \multicolumn{3}{|c}{\textbf{ISCX-17}} 
& \multicolumn{3}{|c}{\textbf{CIC2019}}
\\
\cmidrule[0.8pt](r){1-7}
\textbf{Method}
&{PASS} &{2D-CNN} &{BiLSTN}
&{PASS} &{2D-CNN} &{BiLSTN}

\\
\cmidrule[0.8pt](r){1-7}
{PLI+CPT}
& 83.09 & 70.57 & 68.94 
& 93.19 & 85.14 & 86.65
\\
\cmidrule(r){1-1} \cmidrule(r){2-4} \cmidrule(r){5-7} 
{\emph{w/o} PLI}
& 74.71 & 66.32 & 65.83 
& 90.30 & 83.40 & 85.22
\\
\cmidrule(r){1-1} \cmidrule(r){2-4} \cmidrule(r){5-7}
{\emph{w/o} CPT}
& 78.54 & 68.12 & 67.24 
& 91.24 & 83.17 & 84.81
\\
\cmidrule(r){1-1} \cmidrule(r){2-4} \cmidrule(r){5-7}
{\textcolor{white}{-}\emph{w/o} PLI+CPT\textcolor{white}{-}}
& 73.45 & 63.96 & 62.83 
& 89.31 & 82.56 & 84.18
\\

\bottomrule
\end{tabular}}
\end{table}

\section{Related Work}
\subsection{Encrypted Traffic Classification}
\subsubsection{Fingerprint Construction}
The fingerprints are constructed by extracting plaintext load or unencrypted protocol field, which matched with the fingerprint library by a matching algorithm to identify which the traffic belongs \cite{finsterbusch2013survey}. FlowPrint \cite{van2020flowprint} extracts non-encrypted information to represent each flow and constructs a fingerprint library by clustering and cross-correlation for classification. However, these fingerprints are easily tampered with in communication and lose their correct meaning while not relying on plaintext information.
\subsubsection{Statistical Feature Methods}
Early on, a combination of various traditional machine learning algorithms proposed statistical features to solve ETC tasks. 
Taylor \emph{et al.} \cite{taylor2016appscanner} first designed the burst and flow statistical features and offered a robust app classification method. 
Shi \emph{et al.} \cite{shi2018efficient} built a deep learning framework to select and combine the statistical features to enhance the performance of traffic classification. However, these methods are mainly based on rich experiences, professional knowledge, and much human effort. 

\subsubsection{Sequence Feature Methods}
Encryption apps leak information about dependencies or transfers between data messages, known as sequence features. 
FS-Net \cite{liu2019fs} used packet length or message type sequences as input, joint reconstruction classification loss, with end-to-end training. 
Moreover, Zheng \emph{et al.} \cite{zheng2020learning} proposed a neural network model which using autoencoder to restore the packet length sequence to recognize the app of the flow.
However, this approach relies on vast amounts of labeled data and tends to obtain biased representations of learning minority classes and traffic homogeneity.

\subsection{Class Imbalance}
The class imbalance in the ETC field has not received enough attention, and generic solutions are mainly based on data re-balancing and based on loss re-weighting.

\subsubsection{Based on Data Re-balancing}
The main principle is to balance the number of samples, increase the number of minority classes or reduce the number of the majority through sampling or generation. ROS \cite{japkowicz2000learning} randomly replicates the samples of minority classes, and RUS \cite{drummond2003c4} randomly deletes the samples of majority classes. SMOTE \cite{chawla2002smote} generates data through K-NearestNeighbor samples of features, and the choice of K is related to the number of minority classes. However, data re-balancing requires additional sample generation, which quality will affect classification performance. 

\subsubsection{Based on Loss Re-weighting}
The loss weights of minority classes by re-weighting are made more prominent, thus getting more attention from models. FocalLoss \cite{lin2017focal} improves the traditional cross-entropy loss function and increases the penalty of the loss function for confidence.
FLAGB \cite{guo2020flagb} further combines this approach with the Adaboost algorithm. 
The re-weighting approach is highly applicable and can be changed from the original model, but it reduces effectiveness in data with extreme class imbalance \cite{oksuz2020imbalance}.

\section{Conclusion}
% \section{Conclusion and Future Work}
In this paper, we propose a novel neural network-based ETC method, which can effectively identify massive traffic flows generated by diverse mobile Internet apps. To extract high-quality and robust traffic features, we build a strong backbone feature extractor with a multi-granularity feature construction and multi-head attention encoder. In order to deal with the ubiquitous class imbalance and traffic homogeneity problems in real-world network environments, we optimize the model by contrastive pre-training mechanism. It can improve the model's traffic characterization ability of minority apps and capture more distinguishing features of homogeneous traffic, thereby addressing the label bias issue. In addition, we construct large-scale unlabeled encrypted flows as pseudo-labeled data to perform semi-supervised learning and alleviate the over-reliance on annotated training flows.

We implement our solution as an end-to-end ETC framework, PASS, which shows superior performance compared to state-of-the-art ETC methods and generic sampling approaches. The F1 of PASS achieves 83.09\% to 94.88\% on four datasets with the significant class imbalance and traffic homogeneity issues. Furthermore, extensive ablation studies prove that the multi-granularity features, contrastive pre-training mechanism, and pseudo-label iteration strategy significantly contribute to the performance improvements of PASS. And they can be flexibly migrated to other ETC methods with different feature extractors.

% Although experiments have exhibited the superiority of PASS, there are still many questions to consider. On the one hand, new mobile apps are rapidly emerging on the Internet. In the future, we need to expand PASS to support identifying and collecting new types of encrypted traffic data, making it applicable to a wider range of scenarios. On the other hand, we have observed that when the performance of the initial ETC models is not ideal enough, using pseudo-labeled flows to perform semi-supervised learning may introduce noise into the optimization process, resulting in unstable performance. Furthermore, unlabeled encrypted flows also show significant class imbalance phenomena in many scenarios. In the future, we will further study how to alleviate the noise caused by pseudo-label iteration and exploit unbalanced unlabeled traffic data better.

\end{document}